\documentclass[twocolumn,showpacs,preprintnumbers]{revtex4}
\usepackage{amsmath}
\usepackage{graphicx}
\usepackage{dcolumn}
\usepackage{bm}

\input{tcilatex}
\begin{document}

\title{Dipole-dipole interaction between a quantum dot and graphene nanodisk}
\author{Joel D. Cox$^{1}$, Mahi R. Singh$^{1}$, Godfrey Gumbs$^{2}$, Miguel.
A. Ant\'{o}n$^{3}$ and Fernando Carre\~{n}o$^{3}$}
\affiliation{$^{1}$Department of Physics and Astronomy, The University of Western
Ontario, London, Canada, N6A 3K7}
\affiliation{$^{2}$Department of Physics and Astronomy, Hunter College at the City
University of New York, New\ York 10065, USA}
\affiliation{$^{3}$Escuela Universitaria de Optica, Universidad Complutense de Madrid,
Madrid 28037, Spain}
\affiliation{}

\begin{abstract}
We study theoretically the dipole-dipole interaction and energy transfer in
a hybrid system consisting of a quantum dot and graphene nanodisk embedded
in a nonlinear photonic crystal. In our model a probe laser field is applied
to measure the energy transfer between the quantum dot and graphene nanodisk
while a control field manipulates the energy transfer process. These fields
create excitons in the quantum dot and surface plasmon polaritons in the
graphene nanodisk which interact via the dipole-dipole interaction. Here the
nonlinear photonic crystal acts as a tunable photonic reservoir for the
quantum dot, and is used to control the energy transfer. We have found that
the spectrum of power absorption in the quantum dot has two peaks due to the
creation of two dressed excitons in the presence of the dipole-dipole
interaction. The energy transfer rate spectrum of the graphene nanodisk also
has two peaks due to the absorption of these two dressed excitons.
Additionally, energy transfer between the quantum dot and the graphene
nanodisk can be switched on and off by applying a pump laser to the photonic
crystal or by adjusting the strength of the dipole-dipole interaction. We
show that the intensity and frequencies of the peaks in the energy transfer
rate spectra can be modified by changing the number of graphene monolayers
in the nanodisk or the separation between the quantum dot and graphene. Our
results agree with existing experiments on a qualitative basis. The
principle of our system can be employed to fabricate nano-biosensors,
optical nano-switches, and energy transfer devices.
\end{abstract}

\maketitle

\section{Introduction}

There has been growing interest in developing nanoscale optoelectronic
devices by combining nanomaterials with complementary optical properties
into composite (hybrid) structures. The number of possible composite systems
that can be built from already existing nanostructures is simply enormous. A
significant amount of research on nanocomposites has been devoted to the
study of exciton-plasmon interactions in metal-semiconductor nanostructures,
which offer a wide range of opportunities to control light-matter
interactions and electromagnetic energy flows on nanometer length scales.$%
^{1-6}$ Strong exciton-surface plasmon coupling in semiconductor quantum dot
(QD)-metal nanoparticle systems could lead to efficient transmission of
quantum information between qubits for applications in quantum computing and
communication.$^{2}$ These nanostructures also have applications in
biophotonics and sensing, where nonradiative energy transfer between a QD
and metal nanoparticle can be used to detect biological molecules.$^{3}$

In this paper we study theoretically the dipole-dipole interaction (DDI) and
energy transfer between a quantum emitter and a graphene nanodisk. The
quantum emitter can be a quantum dot, nanocrystal or a chemical or
biological molecule. Here the quantum emitter-graphene system is embedded in
a photonic crystal, which acts as a tunable photonic reservoir for the
emitter. Photonic crystals are engineered, periodically ordered
microstructures that facilitate the trapping and control of light on the
microscopic level. Applications for photonic crystals include all-optical
microchips for optical information processing, optical communication
networks, sensors and solar energy harvesting.$^{7-12}$ In our investigation
we consider a nonlinear photonic crystal, which has a refractive index
distribution that can be tuned optically. The nonlinear photonic crystal
surrounds the hybrid system and is used to manipulate the interaction
between the quantum emitter and graphene nanodisk.

Surface plasmon polaritons are created in the graphene nanodisk due to the
collective oscillations of conduction band electrons. They arise due to the
dielectric contrast between graphene and the surrounding dielectric medium.
Plasmonics is widely studied due to applications in ultrasensitive optical
biosensing,$^{13}$\ photonic metamaterials,$^{14}$\ light harvesting,$^{15}$%
\ optical nanoantennas$^{16}$\ and quantum information processing.$^{17}$
Generally, noble metals are considered as the best available materials for
the study of surface plasmon polaritons.$^{18}$ However, noble metals are
hardly tunable and exhibit large Ohmic losses which limit their
applicability to optical processing devices. Graphene plasmons provide an
attractive alternative to noble-metal plasmons, as they exhibit much tighter
confinement and relatively long propagation distances. Furthermore, surface
plasmons in graphene have the advantage of being highly tunable via
electrostatic gating. Compared to noble metals, graphene also has superior
electronic and mechanical properties, which originate in part from its
charge carriers of zero effective mass.$^{19}$ For example, charge carriers
in graphene can travel for micrometers without scattering at room
temperature. Graphene has also been recognized as a useful optical material
for novel photonic and optoelectronic applications.$^{20-25}$ For these
reasons, the study of plasmonics in graphene has received significant
attention both experimentally and theoretically.$^{21,22,26-29}$

Recently, experimental research on graphene has been extended to the
fabrication and study of QD-graphene nanostructures.$^{30-34}$ For example,
a CdS QD-graphene hybrid system has been synthesized by Cao et al.,$^{30}$
in which a picosecond ultrafast electron transfer process from the excited
QD to the graphene matrix was observed using time-resolved fluorescence
spectroscopy. Chen et al.$^{31}$ have fabricated CdSe/ZnS QDs in contact
with single- and few-layer graphene sheets. By measuring the strong
quenching of the QD fluorescence, they determined the rate of energy
transfer from the QD to graphene. A similar study by Dong et al.$^{32}$ was
performed on a CdTe QD and graphene oxide system, but in their case the QDs
were modified with molecular beacons in order to demonstrate that the hybrid
system can be used for sensing biological molecules. Wang et al.$^{33}$ have
synthesized graphene-CdS and graphene-ZnS QD hybrid systems directly from
graphene oxide, with CdS and ZnS QDs very well dispersed on the graphene
nanosheets. They also measured the QD photoluminescence and observed the
energy transfer between the QDs and graphene. Metal nanoparticle-graphene
hybrid systems have also been fabricated by several groups.$^{21,35-37}$

Here we study a QD-graphene hybrid system, where energy transfer occurs due
to the interaction between optical excitations in the QD and graphene
nanodisk. The optical excitations in the QD are excitons, which are
electron-hole pairs, while those in the graphene nanodisk are surface
plasmon polaritons, which are created due to the collective oscillations of
conduction band electrons. The QD is taken as a three-level lambda-type
system in which two distinct excitonic transitions occur. Other three-level
systems in the ladder-$^{3}$\ and V-type$^{38}$\ configurations interacting
with a metallic nanoparticle in the presence of two external fields have
been studied. In our model we include a probe laser field which is coupled
with one excitonic transition and measures the energy transfer spectra of
the QD and graphene. We also consider that a control laser field is applied
to monitor and control the energy transfer. Aside from creating excitons in
the QD, these fields also generate surface plasmon polaritons in graphene.
The dipoles created by excitons in the QD and plasmons in the graphene
nanodisk then interact via the DDI. This interaction is strong when the QD\
and graphene are in close proximity and their optical excitation frequencies
are resonant.

We have found that the power absorption spectrum of the QD has two peaks
when the QD\ and graphene nanodisk are in close proximity, indicating the
creation of two dressed excitons due to the DDI. These dressed excitons are
transported to graphene, and produce two peaks in the spectrum of the energy
transfer rate to graphene. We show that the energy transfer between the QD
and graphene can be switched on and off by changing the strength of the DDI
coupling or by applying an intense laser field to the nonlinear photonic
crystal. The intensities of peaks in the energy transfer rate spectra can be
controlled by changing the number of graphene monolayers or by changing the
distance between the QD and graphene. We have also predicted that the
intensity of these peaks can be modified in the presence of biological
materials. Our findings agree with the experimental results of Refs. 30-34
on a qualitative basis. The present system can be used to fabricate
nano-biosensors, all-optical nano-switches and energy transfer devices.

\section{Theoretical Formalism}

We investigate theoretically the dipole-dipole interaction\ and energy
transfer between a quantum dot and graphene nanodisk when the system is
embedded in a nonlinear photonic crystal. A schematic diagram for the
present system is shown in Fig. 1. A graphene nanodisk (or nanoflake) lies
in the $x$-$y$ plane, on top of which a QD is deposited. The
center-to-center distance between the QD and graphene nanodisk is taken as $%
R $. The combined QD-graphene nanodisk system can also be referred to as a
QD-graphene nanocomposite or hybrid system. The QD considered here has three
discrete states, where $\left\vert 2\right\rangle $ and $\left\vert
3\right\rangle $ are the lower-energy states which are near-degenerate and
both coupled to the common optically excited state $\left\vert
1\right\rangle $. The QD\ acts as a three-level quantum emitter, where
excitons are created by the transitions $\left\vert 2\right\rangle
\leftrightarrow $ $\left\vert 1\right\rangle $ and $\left\vert
3\right\rangle \leftrightarrow $ $\left\vert 1\right\rangle $ with resonance
frequencies (dipole moments) $\omega _{12}$ ($\mu _{12}$) and $\omega _{13}$
($\mu _{13}$), respectively. This so-called Lambda-type energy level
configuration has been widely studied in atoms, where quantum optical
effects such as electromagnetically induced transparency and coherent
population trapping have been demonstrated.$^{39,40}$ More recently, this
energy level configuration has been achieved in semiconductor QDs.$^{41-43}$%
\FRAME{ftbpFU}{3.3927in}{4.2791in}{0pt}{\Qcb{A schematic diagram of the
QD-graphene nanocomposite embedded in a photonic crystal. The QD has three
discrete states, where $\left\vert 2\right\rangle $ and $\left\vert
3\right\rangle $ denote the lower-energy states, which are near-degenerate
and are both coupled to the common optically excited state $\left\vert
1\right\rangle $.}}{}{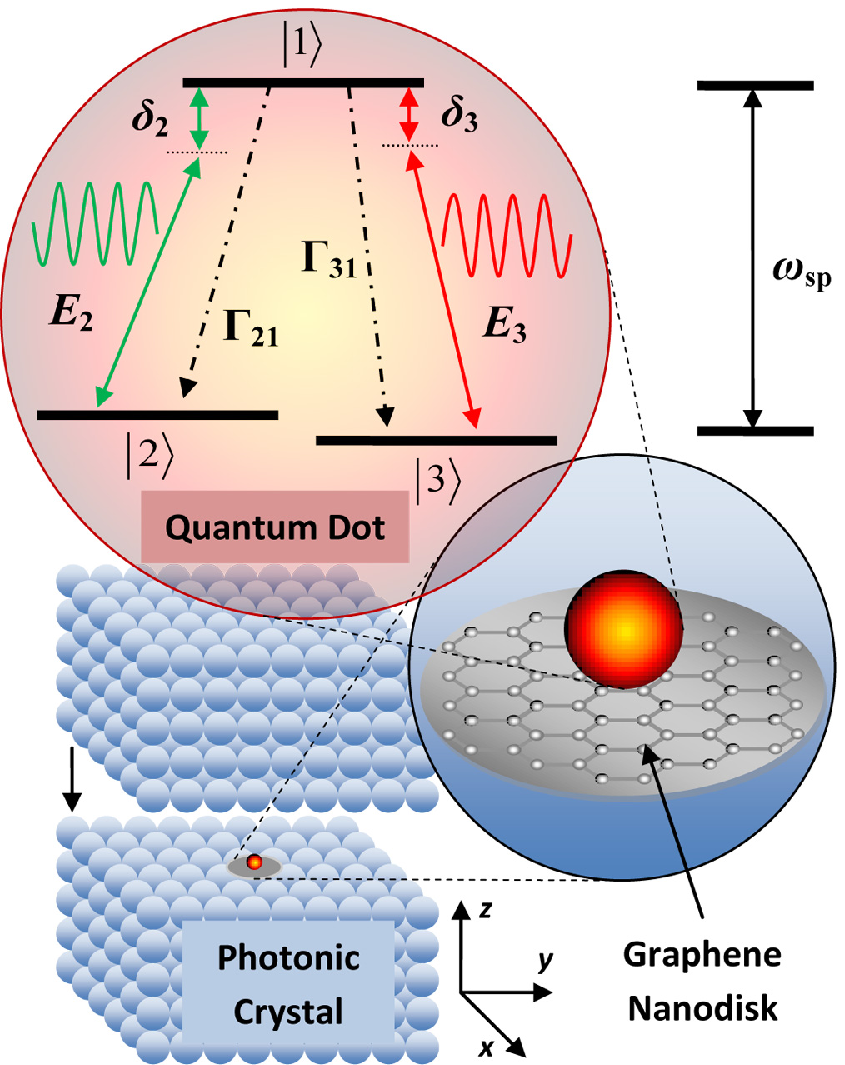}{\special{language "Scientific Word";type
"GRAPHIC";maintain-aspect-ratio TRUE;display "USEDEF";valid_file "F";width
3.3927in;height 4.2791in;depth 0pt;original-width 3.346in;original-height
4.2281in;cropleft "0";croptop "1";cropright "1";cropbottom "0";filename
'figure1.eps';file-properties "XNPEU";}}

In our model, we consider that a probe field $E_{2}=E_{2}^{0}\cos \left(
\omega _{2}t\right) $ is applied between states $\left\vert 1\right\rangle $
and $\left\vert 2\right\rangle $. To control the coupling between the QD and
graphene a control field $E_{3}=E_{3}^{0}\cos \left( \omega _{3}t\right) $
is applied between states $\left\vert 1\right\rangle $ and $\left\vert
3\right\rangle $. These laser fields excite both the QD and graphene
nanodisk. In the\ QD, the probe and control fields create excitons which
produce dipole electric fields that interact with the graphene nanodisk.
Similarly in the graphene nanodisk surface plasmons are generated by the
probe and control fields, which produce dipole electric fields that interact
with the QD.

The dipole electric fields produced by the QD and graphene are written as%
\begin{eqnarray}
E_{DDI}^{QD} &=&\frac{g_{l}P_{QD}}{4\pi \epsilon _{b}R^{3}}  \TCItag{1}
\label{E_QD} \\
E_{DDI}^{G} &=&\frac{g_{l}P_{G}}{4\pi \epsilon _{b}R^{3}}  \notag
\end{eqnarray}%
respectively. Here $\epsilon _{bd}=\left( 2\epsilon _{b}+\epsilon
_{d}\right) /3\epsilon _{b}$, where $\epsilon _{d}$ is the dielectric
constant of the QD and $\epsilon _{b}$ is the dielectric constant of the
medium surrounding the QD-graphene system. The parameter $g_{l}$ ($l=x$, $y$%
, $z$) is the polarization parameter, with $g_{x}=g_{y}=-1$ and $g_{z}=2$
for electric fields polarized in the $x$-$y$ plane or in the $z$-direction,
respectively.$^{44}$ In Eq. (1) $P_{QD}$ and $P_{G}$ are the polarization of
the quantum dot and graphene nanodisk, respectively. The polarization of the
QD is obtained as

\begin{equation}
P_{QD}=\sum_{i=2,3}\mu _{1i}\left( \rho _{1i}+\rho _{i1}\right)  \tag{2}
\end{equation}%
where $\mu _{ij}$ and $\rho _{ij}$ are the dipole moment and density matrix
element, respectively, for the transition $\left\vert i\right\rangle
\leftrightarrow \left\vert j\right\rangle $.

The total electric field felt by the graphene nanodisk is written as 
\begin{equation}
E_{G}=E_{2}+E_{3}+E_{DDI}^{QD}  \tag{3}
\end{equation}%
where the first, second and third terms are the contributions from the
probe, control and the QD dipole field, respectively. Using the quasistatic
dipole approximation,$^{44}$ the polarization in the graphene nanodisk is 
\begin{eqnarray}
P_{G} &=&\epsilon _{b}\alpha _{l}\left( E_{2}+E_{3}+E_{DDI}^{QD}\right) 
\TCItag{4} \\
\alpha _{l} &=&\frac{4\pi L_{x}L_{y}L_{z}\left[ \epsilon _{m}\left( \omega
\right) -\epsilon _{b}\right] }{3\epsilon _{b}+3\varsigma _{l}\left[
\epsilon _{m}\left( \omega \right) -\epsilon _{b}\right] }  \TCItag{5}
\end{eqnarray}%
where $\alpha _{l}$ is the polarizability of the graphene nanodisk and $%
\epsilon _{g}\left( \omega \right) $ denotes the dielectric function of
graphene. Here $\varsigma _{l}$ is called the depolarization factor, and is
obtained as%
\begin{equation}
\varsigma _{l}=\frac{L_{x}L_{y}L_{z}}{2}\int_{0}^{\infty }\frac{dq}{\left(
L_{l}^{2}+q\right) f\left( q\right) }  \tag{6}
\end{equation}%
where%
\begin{equation}
f\left( q\right) =\sqrt{\left( L_{x}^{2}+q\right) \left( L_{y}^{2}+q\right)
\left( L_{z}^{2}+q\right) }  \tag{7}
\end{equation}%
The depolarization parameters satisfy the relation $\varsigma _{x}+\varsigma
_{y}+\varsigma _{z}=1$, and determine the optical response of the graphene
nanodisk based on its shape.$^{44}$ For an oblate spheroid where $%
L_{x}=L_{y} $ and $L_{x}>L_{z}$, the depolarization factors reduce to%
\begin{eqnarray}
\varsigma _{z} &=&\frac{1-e_{g}^{2}}{e_{g}^{2}}\left[ \frac{1}{2e_{g}}\log
\left( \frac{1+e_{g}}{1-e_{g}}\right) -1\right]  \TCItag{8} \\
\varsigma _{x} &=&\varsigma _{y}=\frac{1}{2}\left( 1-\varsigma _{z}\right) 
\notag
\end{eqnarray}%
where the eccentricity of the nanodisk is defined as $e_{g}=\sqrt{1-\left(
L_{x}/L_{z}\right) ^{2}}$. Note that this equation is valid for both cases
where $L_{x}>L_{z}$ and $L_{x}<L_{z}$.$^{44}$

The above expression for the polarizability given in Eq. (5) has been widely
used in the literature to study the optical properties of metallic
nanodisks, and has been found to give good agreement with experimental
results.$^{45,46}$ In the quasistatic approximation the dimensions of the
graphene nanodisk are much smaller than the wavelength of incident light and
we therefore assume a spatially uniform but time-varying electric field
across the graphene nanodisk. Here the wavelengths of light considered are
on the order of several hundred nanometers, and thus the size of the
graphene nanodisk must be less than $100$ nm. It is important to note that
our model is only valid for nano-size graphene samples and not for bulk
materials. For a very flat and thin disk we take $L_{x}\gg L_{z}$, which
gives$^{47,48}$ 
\begin{eqnarray}
\varsigma _{x} &=&\varsigma _{y}\cong \frac{\pi }{4}\frac{L_{z}}{L_{x}} 
\TCItag{9} \\
\varsigma _{z} &\cong &1-\frac{\pi }{2}\frac{L_{z}}{L_{x}}  \notag
\end{eqnarray}%
The above method has been used for a graphene\ flake in Ref. [49].

The total electric field felt by the QD is written as%
\begin{equation}
E_{QD}=\frac{E_{2}}{\epsilon _{bd}}+\frac{E_{3}}{\epsilon _{bd}}+\frac{%
g_{l}\alpha _{l}\left( E_{2}+E_{3}+E_{DDI}^{QD}\right) }{4\pi \epsilon
_{bd}R^{3}}  \tag{10}
\end{equation}%
where the first, second and third terms are the contributions from the
probe, control and the graphene dipole field, respectively. Putting the
expression of $E_{DDI}^{QD}$ from Eq. (1) into the above expression we get%
\begin{eqnarray}
E_{QD} &=&\sum_{i=2,3}\left( E_{F}^{i}+E_{DDI}^{i}\right)  \TCItag{11} \\
E_{F}^{i} &=&\frac{\hbar }{\mu _{1i}}\Omega _{i}e^{-i\omega _{i}t}+c.c. 
\notag \\
E_{DDI}^{i} &=&\frac{\hbar }{\mu _{1i}}\left( \Pi _{i}+\Lambda _{i}\rho
_{1i}\right) e^{-i\omega _{i}t}+c.c.  \notag
\end{eqnarray}%
where 
\begin{eqnarray}
\Omega _{i} &=&\frac{\mu _{1i}E_{i}^{0}}{2\hbar \epsilon _{bd}}  \TCItag{12}
\\
\Pi _{i} &=&\frac{g_{l}\alpha _{l}\left( \omega _{i}\right) \Omega _{i}}{%
4\pi R^{3}}  \notag \\
\Lambda _{i} &=&\frac{g_{l}^{2}\alpha _{l}\left( \omega _{i}\right) \mu
_{1i}^{2}}{\left( 4\pi \right) ^{2}\epsilon _{b}\hbar \epsilon _{bd}^{2}R^{6}%
}  \notag
\end{eqnarray}

The dipoles of the QD interact with the dipole electric field produced by
the graphene nanodisk, and vice versa. This interaction is called the
dipole-dipole interaction (DDI). The terms in the interaction Hamiltonian of
the QD due to the external probe and control fields and the DDI are
expressed in the rotating wave approximation as%
\begin{eqnarray}
H_{QD-F} &=&-\sum_{i=2,3}\mu _{1i}E_{F}^{i}\sigma _{1i}^{+}+h.c.  \TCItag{13}
\\
H_{QD-DDI} &=&-\sum_{i=2,3}\mu _{1i}E_{DDI}^{i}\sigma _{1i}^{+}+h.c.  \notag
\end{eqnarray}%
Using the expressions for $E_{F}^{i}$\ and $E_{DDI}^{i}$\ in Eqs. (11) in
the above expression and putting the Hamiltonian in interaction
representation we get 
\begin{eqnarray}
H_{QD-F} &=&-\sum_{i=2,3}\hbar \Omega _{i}\sigma _{1i}^{+}e^{-i\left( \omega
_{i}-\omega _{1i}\right) t}+h.c.  \TCItag{14}  \label{H_F} \\
H_{QD-DDI} &=&-\sum_{i=2,3}\hbar \left( \Pi _{i}+\text{\ }\Lambda _{i}\rho
_{1i}\right) \sigma _{1i}^{+}e^{-i\left( \omega _{i}-\omega _{1i}\right)
t}+h.c.  \notag
\end{eqnarray}%
where $\sigma _{ij}^{+}=\left\vert i\right\rangle \left\langle j\right\vert $
($\sigma _{ij}=\left\vert j\right\rangle \left\langle i\right\vert $) is the
exciton creation\ (annihilation) operator. The interaction Hamiltonian term $%
H_{QD-F}$ given in Eq. (14) represents the direct contribution from the
external probe ($i=2$) and control ($i=3$) fields incident on the QD, and is
given in terms of the Rabi frequencies $\Omega _{i}$ which measure the
intensities of these fields. The second contribution $H_{QD-DDI}$ represents
the fields incident on the QD due to the DDI between the\ QD\ and graphene
nanodisk, and contains two terms $\Pi _{i}$ and $\Lambda _{i}\rho _{1i}$.
The first DDI term is due to the interaction of the QD with the dipole
electric fields from the graphene nanodisk induced by the probe and control
fields. Therefore we refer to this as the \textit{direct DDI}\ term. The
second DDI contribution is due to the interaction of the QD with a dipole
field from graphene\ that arises when the external fields polarize the QD,
which in turn polarize graphene. In other words these contributions are the
self-interaction of the QD, as they depend on the polarization of the QD.
For this reason this term is called the \textit{self-induced DDI} parameter.
The surface plasmon polariton resonance frequency $\omega _{sp}^{l}$ in the
graphene nanodisk is obtained by setting the real part of the denominator in 
$\alpha _{l}\left( \omega \right) $ equal to zero and solving for $\omega $.
When the optical excitation frequencies of the QD lie near the surface
plasmon polariton resonance frequencies of the graphene nanodisk (i.e. $%
\omega _{1i}\approx \omega _{sp}$), the DDI becomes very strong due to the
enhanced local fields in the vicinity of the graphene nanodisk. This
interaction leads to excitation and energy transfer between the QD and
graphene.

The combined QD-graphene system is embedded in a photonic crystal consisting
of dielectric spheres arranged periodically in three dimensions, which acts
as a reservoir for the QD.\ Therefore, we consider that the excited state $%
\left\vert 1\right\rangle $ spontaneously decays to the lower-energy states $%
\left\vert 2\right\rangle $ and $\left\vert 3\right\rangle $ due to excitons
coupling with Bloch photons in the photonic crystal (see Fig. 1). In this
case the interaction Hamiltonian is given as:%
\begin{align}
H_{QD-PC}& =-\sum_{i=2,3}\sum_{k}g_{1i}^{PC}a_{k}\sigma _{1i}^{+}e^{i\left(
\omega _{1i}-\omega _{k}\right) t}+h.c.\text{,}  \tag{15}  \label{H_PC} \\
g_{1i}^{PC}& =\sqrt{\frac{\hbar \omega _{k}}{2\epsilon _{b}V_{PC}}}\left( 
\mathbf{e}_{k}\cdot \mathbf{\mu }_{1i}\right)  \notag
\end{align}%
where $\mathbf{e}_{k}$ is the polarization unit vector of the Bloch photons
in the PC and $V_{PC}$ is the volume of the photonic crystal. The operator $%
a_{k}^{+}$ ($a_{k}$) is the photon creation (annihilation) operator, while $%
\omega _{k}$ and $k$ are the Bloch photon frequency and wave vector,
respectively.

Finally, the total interaction Hamiltonian of the system is written as%
\begin{eqnarray}
H &=&H_{QD-F}+H_{QD-DDI}+H_{QD-PC}  \TCItag{16}  \label{H_0} \\
&=&-\sum_{i=2,3}\hbar \left( R_{i}e^{i\theta _{i}}+\Lambda _{i}\rho
_{1i}\right) \sigma _{1i}^{+}e^{-i\left( \omega _{i}-\omega _{1i}\right) t} 
\notag \\
&&-\sum_{i=2,3}\sum_{k}g_{1i}^{PC}a_{k}\sigma _{1i}^{+}e^{i\left( \omega
_{1i}-\omega _{k}\right) t}+h.c.  \notag
\end{eqnarray}%
where%
\begin{align*}
R_{i}& =\sqrt{\left[ \Omega _{i}+\func{Re}(\Pi _{i})\right] ^{2}+\left[ 
\func{Im}(\Pi _{i})\right] ^{2}} \\
\theta _{i}& =\arctan \left[ \frac{\func{Im}(\Pi _{i})}{\Omega _{i}+\func{Re}%
(\Pi _{i})}\right]
\end{align*}%
Note that the first term in the above expression for $H$ is the combination
of $H_{QD-F}$ with the direct DDI term.

We use the density matrix method to evaluate the energy transfer between the
QD and the graphene. Using Eq. (16) for the interaction Hamiltonian and the
master equation for the density matrix$^{39}$ we obtained the following
equations of motion for the QD density matrix elements as%
\begin{align}
\frac{d\rho _{22}}{dt}& =2\Gamma _{21}\rho _{11}-iR_{2}e^{i\theta _{2}}\rho
_{21}-i\Lambda _{2}\rho _{12}\rho _{21}  \tag{17}  \label{rho_first} \\
& +iR_{2}e^{-i\theta _{2}}\rho _{12}+i\Lambda _{2}^{\ast }\rho _{21}\rho
_{12}  \notag \\
\frac{d\rho _{33}}{dt}& =2\Gamma _{31}\rho _{11}-iR_{3}e^{i\theta _{3}}\rho
_{31}-i\Lambda _{3}\rho _{13}\rho _{31}  \tag{18} \\
& +iR_{3}e^{-i\theta _{3}}\rho _{13}+i\Lambda _{3}^{\ast }\rho _{31}\rho
_{13}  \notag \\
\frac{d\rho _{12}}{dt}& =-d_{12}\rho _{12}+iR_{3}e^{i\theta _{3}}\rho
_{32}+i\Lambda _{3}\rho _{13}\rho _{32}  \tag{19}  \label{rho_12} \\
& -iR_{2}e^{-i\theta _{2}}(\rho _{11}-\rho _{22})  \notag \\
\frac{d\rho _{13}}{dt}& =-d_{13}\rho _{13}+iR_{2}e^{i\theta _{2}}\rho
_{23}+i\Lambda _{2}\rho _{12}\rho _{23}  \tag{20} \\
& -iR_{3}e^{i\theta _{3}}(\rho _{11}-\rho _{33})  \notag \\
\frac{d\rho _{23}}{dt}& =-i(\delta _{2}-\delta _{3})\rho
_{23}+iR_{2}e^{-i\theta _{2}}\rho _{13}+i\Lambda _{2}^{\ast }\rho _{21}\rho
_{31}  \tag{21}  \label{rho_last} \\
& -iR_{3}e^{i\theta _{3}}\rho _{21}-i\Lambda _{3}\rho _{13}\rho _{21}  \notag
\end{align}%
where%
\begin{eqnarray*}
d_{1i} &=&\Gamma _{21}+\Gamma _{31}-\Gamma _{id}-i\Delta _{id}-i\delta _{i}%
\text{,} \\
\delta _{i} &=&\omega _{i}-\omega _{1i}\text{.}
\end{eqnarray*}%
Here $\delta _{i}$ are the detuning of the probe ($i=2$) and control ($i=3$)
fields. Note that the diagonal elements of the density matrix satisfy the
relation $\rho _{11}+\rho _{22}+\rho _{33}=1$. The quantities $\Gamma _{id}$
and $\Delta _{id}$ are the \textit{non-radiative decay rate} and \textit{%
energy shift}, respectively, due to self-induced DDI parameters $\Lambda
_{i} $. They are found as%
\begin{align*}
\ \Gamma _{id}& =\func{Im}(\Lambda _{i})\left( \rho _{ii}-\rho _{11}\right)
\\
\Delta _{id}& =\func{Re}(\Lambda _{i})\left( \rho _{ii}-\rho _{11}\right)
\end{align*}%
The parameters $\Gamma _{i1}$ represent the spontaneous decay rates of
excited state $\left\vert 1\right\rangle $ to state $\left\vert
i\right\rangle $ due to the Bloch photons in the photonic crystal, and are
given as%
\begin{equation}
\Gamma _{i1}=\Gamma _{i1}^{0}\frac{\pi ^{2}c^{3}}{V_{PC}\omega _{1i}^{2}}%
D(\omega _{1i})  \tag{22}
\end{equation}%
where%
\begin{eqnarray*}
D(\omega ) &=&\sum\limits_{\pm }\frac{\xi _{\pm }^{0}(\omega )V_{PC}\arccos
^{2}\left[ F(\omega )\right] }{\sqrt{1-F^{2}(\omega )}}\text{,} \\
\xi _{\pm }^{0}(\omega ) &=&\frac{\left( n_{a}\pm n_{b}\right) ^{2}\left(
n_{a}a\pm n_{b}b\right) }{2\pi ^{2}L^{3}cn_{a}n_{b}}\sin \left[ \frac{%
2\omega }{c}\left( n_{a}a\pm n_{b}b\right) \right] \text{,}
\end{eqnarray*}%
and%
\begin{equation*}
F(\omega )=\sum_{\pm }\pm \frac{\left( n_{a}\pm n_{b}\right) ^{2}}{%
4n_{a}n_{b}}\cos \left[ \frac{2\omega }{c}\left( n_{a}a\pm n_{b}b\right) %
\right] \text{.}
\end{equation*}%
In the above expression, $\Gamma _{i1}^{0}$ is the exciton\ decay rate due
to the background radiation field in free space. Here $L=2a+2b$ is the
photonic crystal lattice constant, $2b$ is the spacing between dielectric
spheres and $a$ is the radius of the spheres. Parameters $n_{a}$ and $n_{b}$
denote the refractive index of the dielectric spheres and background
material in the photonic crystal, respectively. The expression for the
photonic density of states $D(\omega )$ has been derived in Ref. [50]. Here
we have used the Markovian approximation in order to derive the decay rates
for the QD in the presence of the photonic crystal. This approximation
ignores memory effects in the electromagnetic reservoir due to the presence
of the QD, and is valid when the photonic density of states\ can be
considered smooth and slowly-varying compared to the energy difference
between the edge of the photonic band gap and the resonance frequency of the
QD.$^{51}$ Note that in our calculations, we remain within the regime where
the Markovian approximation is valid. Therefore the effect of the photonic
crystal serves only to alter the decay rates of the excitonic transitions
compared to those in free space. We numerically solve Eqs. (17--21) by first
substituting $\rho _{12}=\widetilde{\rho }_{12}e^{i\theta _{2}}$, $\rho
_{13}=\widetilde{\rho }_{13}e^{i\theta _{3}}$ and $\rho _{23}=\widetilde{%
\rho }_{23}e^{-i\left( \theta _{2}-\theta _{3}\right) }$.

Following the method of Ref. [44] and using Eq. (16), the energy absorption
rate of the QD ($W_{QD}$) and the energy transfer rate from the QD to
graphene ($W_{G}$) are found as 
\begin{align}
W_{QD}& =\sum_{i=2,3}\hbar \omega _{1i}\rho _{11}\Gamma _{i1}  \tag{23}
\label{W_QD} \\
W_{G}& =\sum_{i=2,3}\frac{g_{x,z}^{2}\mu _{1i}^{2}\omega _{i}\func{Im}%
(\alpha _{x})\left\vert \widetilde{\rho }_{1i}\right\vert ^{2}}{8\pi
^{2}\epsilon _{b}\epsilon _{bd}^{2}\left\vert \epsilon _{bg}\right\vert
^{2}R^{6}}  \tag{24}  \label{W_G}
\end{align}%
where $\epsilon _{bg}=\left( 2\epsilon _{b}+\epsilon _{g}\right) /3\epsilon
_{b}$. The expression for $W_{QD}$ is obtained by assuming that the power
radiated from the QD is equal to its energy absorption rate. Similar
expressions have been widely used in the literature on hybrid systems.$%
^{2,3} $ Note that the energy transfer to graphene depends on the coherences 
$\widetilde{\rho }_{1i}$ of the QD density matrix, which change depending on
the center-to-center distance between the QD and graphene, $R$. Therefore
Eq. (24) does not simply depend on $R^{6}$, but rather is a much more
complicated function of $R$.

\section{Results and Discussion}

We consider a graphene nanodisk with thickness $L_{z}=0.35$ nm (i.e. a
single graphene layer) and size ratio $L_{x}/L_{z}$ $=20$. The plasmon
frequency and background dielectric constant of graphene are taken from
experiments as $6.02$ eV and $1.964$, respectively.$^{21}$ The decay rate in
graphene is taken as $\gamma _{G}=5$ THz, which is consistent with the
relaxation rates reported in refs. [22] and [26]. With these parameters, the
surface plasmon polariton resonance frequencies in the graphene\ nanodisk
are calculated as $\hbar \omega _{sp}^{x}=0.8026$ eV and $\hbar \omega
_{sp}^{z}=4.1250$ eV. The QD dielectric constant and dipole moments are
taken as $\epsilon _{d}=12$ and $\mu _{12}=\mu _{13}=0.1$ $e$ $\times $ nm,
respectively, while the free space decay rates for the QD\ are taken as $%
\Gamma _{21}^{0}=\Gamma _{31}^{0}=0.2$ $\mu $eV. These parameters are
comparable to those commonly found in the literature for QDs.$^{2-4}$ Here
the transition energies in the QD are taken to lie near the plasmon
resonance $\hbar \omega _{sp}^{x}$ as $\hbar \omega _{12}=0.8046$ eV and $%
\hbar \omega _{13}=0.8036$ eV. The combined QD-graphene nanodisk hybrid is
contained within a photonic crystal made of polystyrene spheres arranged
periodically in air. Similar photonic crystals have been fabricated by Liu
et al.$^{9}$, in which ultrafast all-optical switching was experimentally
demonstrated. Photonic crystal parameters are taken as $a=170$ nm, $L=480$
nm, $n_{a}=1.59$ and $n_{b}=1$. With these parameters, we find that a
photonic band gap appears between frequencies $0.8225$ eV and $0.9843$ eV.
Note that the lower edge of the band gap lies near $\omega _{sp}^{x}$ and
the QD transition frequencies $\omega _{12}$ and $\omega _{13}$. The vacuum
decay rates for the QD\ are taken as $\Gamma _{2}^{0}=\Gamma _{3}^{0}=0.2$ $%
\mu $eV, and in the presence of the photonic crystal it is found that $%
\Gamma _{21}=1.1370$ $\mu $eV and $\Gamma _{31}=1.1127$ $\mu $eV. Here the
background dielectric constant was taken as $\epsilon _{B}=2.081$.
Throughout the following calculations we consider that the intensity of the
probe and control fields are $1.0$ and $3.0$ W/cm$^{2}$, respectively.

We first consider the case where the excitonic transition $\left\vert
3\right\rangle \leftrightarrow \left\vert 1\right\rangle $ is coupled with
the surface plasmon resonance of the graphene nanodisk. In this
configuration, the transition frequency $\omega _{13}$ is near $\omega
_{sp}^{x}$ while both the control field $E_{3}$ and the transition dipole
moment $\mu _{13}$ are polarized in the $x$-$y$ plane. Conversely, the
transition $\left\vert 2\right\rangle \leftrightarrow \left\vert
1\right\rangle $ is not coupled with the graphene nanodisk. This situation
occurs when the probe field $E_{2}$ and transition dipole moment $\mu _{12}$
are polarized in the $z$-direction and $\omega _{12}$ is far away from $%
\omega _{sp}^{z}$ (see Fig. 2 inset). The energy absorption rate in the QD
is evaluated from Eq. (23) and the results are presented in Fig. 2a when the
QD-graphene separation $R$ is varied and the control field is resonant with
the $\left\vert 3\right\rangle \leftrightarrow \left\vert 1\right\rangle $
transition such that $\delta _{3}=0$. It is found that the power absorption
spectrum has a single peak with an extremely narrow transparent window at $%
\delta _{2}=0$ when the QD and graphene are further away with each other
(i.e. $R=20$ nm). This narrow minima is due to electromagnetically induced
transparency in the system. When the QD is close to graphene (i.e. $R=8$ nm)
the power absorption peak splits into two peaks and a clear minima appears
at $\delta _{2}=0$. The observed splitting is due to the DDI and surface
plasmon polariton coupling.\FRAME{ftbpFU}{3.3892in}{3.7844in}{0pt}{\Qcb{%
Energy absorption rate of the QD\ as a function of probe field detuning $%
\protect\delta _{2}$ when the QD-graphene nanodisk\ separation $R$ is
varied. (a) $\protect\delta _{3}=0$; (b) $\protect\delta _{3}=10$ $\protect%
\mu $eV. Inset: Polarization of the probe and control fields.}}{}{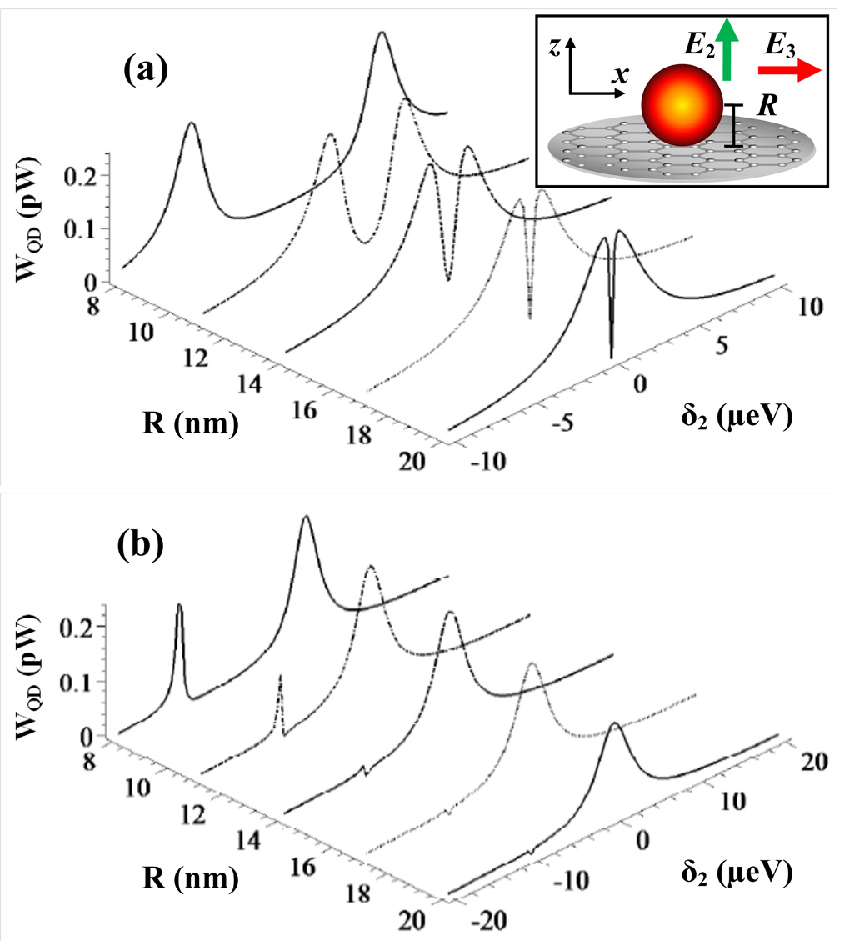%
}{\special{language "Scientific Word";type "GRAPHIC";maintain-aspect-ratio
TRUE;display "USEDEF";valid_file "F";width 3.3892in;height 3.7844in;depth
0pt;original-width 3.3434in;original-height 3.7369in;cropleft "0";croptop
"1";cropright "1";cropbottom "0";filename 'figure2.eps';file-properties
"XNPEU";}}

The splitting of the power absorption spectrum can be explained using the
theory of dressed states. When the QD is close to the graphene nanodisk
there is strong coupling due\ to the DDI for the transition $\left\vert
3\right\rangle \leftrightarrow \left\vert 1\right\rangle $. This causes the
excited state $\left\vert 1\right\rangle $ to split into two dressed states,
namely $\left\vert 1_{+}\right\rangle $ and $\left\vert 1_{-}\right\rangle $%
. Therefore, there are now two transitions $\left\vert 2\right\rangle
\leftrightarrow \left\vert 1_{+}\right\rangle $ and $\left\vert
2\right\rangle \leftrightarrow \left\vert 1_{-}\right\rangle $ which give
two peaks and and a minima in the spectrum. In other words, a single exciton
splits into two dressed excitons, and their energy difference is found to be
proportional to the DDI. As the distance between the QD and graphene
increases, the splitting decreases since the direct DDI term $\Pi _{i}$ is
inversely proportional to $R^{3}$. In Fig. 2b we have plotted the energy
absorption rate when the control field is detuned such that $\delta _{3}=10$ 
$\mu $eV. In this case, we find one peak and negligible
electromagnetically-induced transparency\ when $R$ is large. When $R$
decreases, the single peak splits into two peaks due to the DDI. These
results show that the DDI\ can be used to split one exciton into two
excitons, and also to control the electromagnetically-induced transparency
phenomenon.

In Fig. 3 we have investigated the effect of the photonic crystal on the
energy absorption rate in the lambda-type QD. Initially, the lower band edge
of the photonic crystal lies far away from the resonance energies of the QD
(see solid curve), and there is weak coupling between the QD and photonic
crystal. When we move the lower band edge closer to the resonance frequency $%
\omega _{13}$ of the QD, the two peaks in the power absorption spectrum
merge into a broad peak with a narrow electromagnetically-induced
transparency window at $\delta _{2}=0$ (see dashed curve). Note also that
the height of the peaks decreases. The merging of the split peaks in the QD
power absorption spectrum occurs because the spontaneous decay rates become
larger than\ the DDI splitting for the two peaks. The value of the decay
rate is large because the photonic density of states is large when the
resonance energy of the QD lies near the band edges. For example, we found $%
\Gamma _{21}=6.40\ \mu $eV and $\Gamma _{31}=3.81$ $\mu $eV whereas the
energy splitting is about $2.80$ $\mu $eV. Here, the location of the
photonic crystal\ band edges can be changed by applying an intense pulsed
laser field. The intense laser field causes the refractive index of
polystyrene, an optical nonlinear material, to change due to the Kerr
effect. This change is quantified by the expression $n_{a}^{\prime
}=n_{a}+n_{3}I_{pump}$ where $n_{3}$ is the Kerr nonlinearity constant and
has the value $n_{3}=1.15\times 10^{-12}$ cm$^{2}$/W for polystyrene.$^{10}$
For the pump field intensity $I_{pump}=31.0$ GW/cm$^{2}$ we found that the
photonic crystal\ band edge shifts such that $\Delta \varepsilon _{v}=-$ $%
17.32$ meV. This means that the hybrid system can be used to study the
nonlinear properties of photonic crystals. Using an external pump field to
induce a large Kerr nonlinearity in the polystyrene photonic crystal is also
an effective way to switch the energy transfer between two states; from high
to low energy transfer peaks. Alternatively, the refractive index of the
background material in the photonic crystal can also be modified by
immersing the photonic crystal in another material. Therefore the present
QD-graphene system can also be used as a nano-sensor.\FRAME{ftbpFU}{3.3918in%
}{3.2932in}{0pt}{\Qcb{Energy absorption rate in the QD as a function of
probe field detuning $\protect\delta _{2}$ when the lower band edge of the
photonic crystal is taken as $\protect\varepsilon _{v}=\hbar \protect\omega %
_{12}+17.88$ meV (a) and $\protect\varepsilon _{v}=\hbar \protect\omega %
_{12}+0.56$ meV (b). Here $R=13$ nm and $\protect\delta _{3}=0$. Inset:
Polarization of the probe and control fields.}}{}{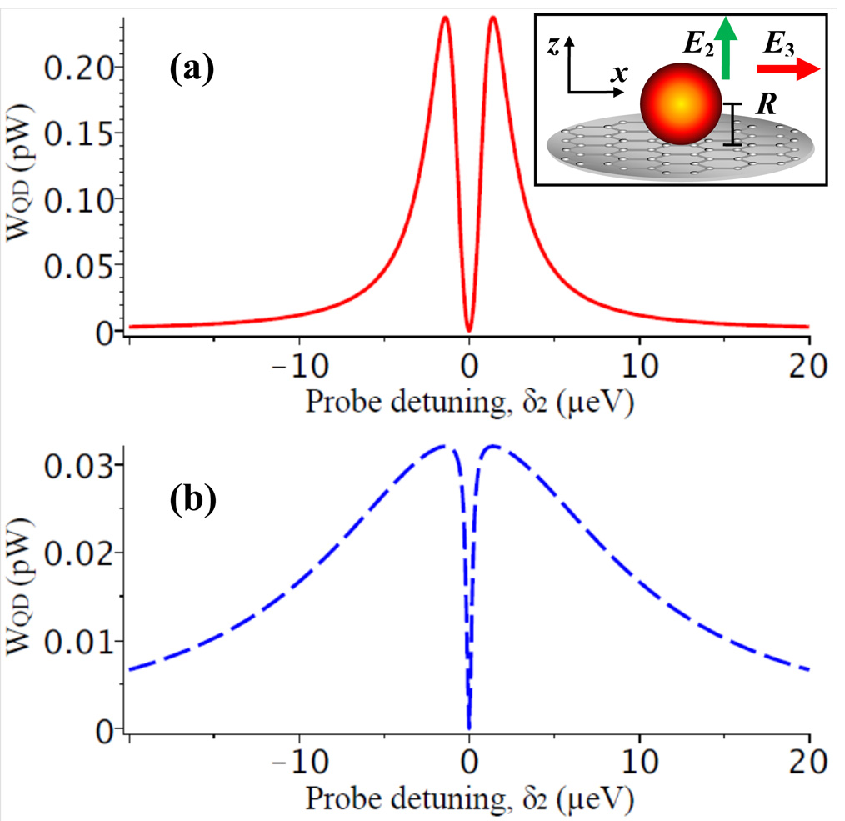}{\special%
{language "Scientific Word";type "GRAPHIC";maintain-aspect-ratio
TRUE;display "USEDEF";valid_file "F";width 3.3918in;height 3.2932in;depth
0pt;original-width 3.3451in;original-height 3.2482in;cropleft "0";croptop
"1";cropright "1";cropbottom "0";filename 'figure3.eps';file-properties
"XNPEU";}}

We now investigate the energy transfer rate from the QD to graphene, and the
results are plotted in Fig. 4 as a function of probe detuning when the
QD-graphene\ separation $R$ is varied. We find that the power transfer
spectrum has a single peak with a narrow electromagnetically-induced
transparency\ window when $R$ is large (i.e. $R=20$ nm). When the\ QD is
brought closer to graphene\ (i.e. $R=8$ nm), the power transfer spectrum has
one large minima and two peaks with separation proportional to the DDI. This
indicates that energy is transferred from the QD to the graphene when the
two dressed excitons created in the QD are absorbed by graphene. This is an
interesting finding, and can be used to transfer energy absorbed by the QD\
from a light source (i.e. the sun) to graphene where it can be stored.
Therefore, one can fabricate energy transfer and storage devices (i.e. solar
cells) from the present hybrid system.\FRAME{ftbpFU}{3.3918in}{2.271in}{0pt}{%
\Qcb{Energy transfer\ rate from the QD to graphene\ as a function of probe
field detuning $\protect\delta _{2}$ when the QD-graphene nanodisk\
separation $R$ is varied and $\protect\delta _{3}=0$. Inset: Polarization of
the probe and control fields.}}{}{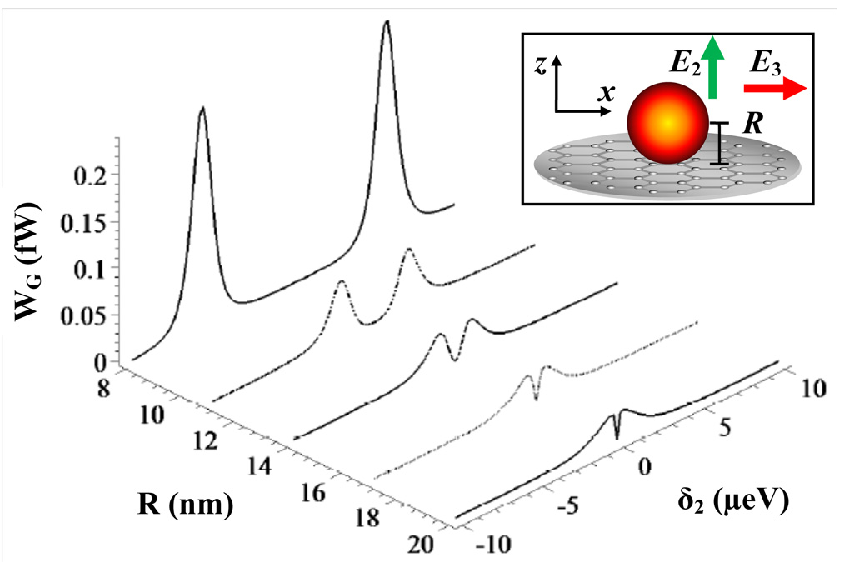}{\special{language "Scientific
Word";type "GRAPHIC";maintain-aspect-ratio TRUE;display "USEDEF";valid_file
"F";width 3.3918in;height 2.271in;depth 0pt;original-width
3.3451in;original-height 2.2312in;cropleft "0";croptop "1";cropright
"1";cropbottom "0";filename 'figure4.eps';file-properties "XNPEU";}}

It is also found that the height of the energy transfer peaks increases as
the QD-graphene separation decreases (see Fig. 4). This effect has been
observed experimentally by Chen et al.$^{31}$ and Dong et al.$^{32}$ They
found that as the distance between CdTe-QDs and a graphene oxide sheet
decreases there is a strong quenching of the QD fluorescence. They concluded
that the fluorescence quenching could be due the energy transfer from the QD
to the graphene sheet. For example, Chen et al.$^{31}$ deposited graphene on
quartz substrates and then CdSe/ZnS-QDs were deposited on graphene. The
fluorescence measurements were performed on the individual QDs located both
on the bare quartz substrate and on a graphene layer. They observed strong
fluorescence quenching for QDs deposited on the graphene sheet, which was
attributed to the energy transfer between QD and graphene and not due to
photoinduced electron transfer from the QD to graphene. Similarly Wang et al.%
$^{33}$ performed photoluminescence measurements on CdS--QDs and ZnS--QDs on
graphene and observed a strong quenching of photoluminescence for these QDs
due to the presence of the graphene sheet. They also performed transient
photovoltaic experiments on their hybrid systems and found a very unexpected
strong positive photovoltaic response due to the DDI. Conversely, it was
found that separate samples of graphene and CdS-QDs of a similar size do not
show any photovoltaic response.\ They concluded their experimental findings
can be explained due to the energy transfer between the QD and the graphene
sheet.\ Similar energy transfer between a QD and carbon nanotube has also
been found experimentally by Shafran et al..$^{34}$

When a QD is in contact with biomolecules, molecular beacons, DNA or
aptamers, its dielectric constant can be modified. Therefore we have
investigated the role of the dielectric constant of the QD on the energy
transfer to graphene. The results are plotted in Fig. 5a for three values of 
$\epsilon _{d}$. It is found that by changing the dielectric constant of the
QD, the height of the energy transfer peaks can be modified. For example by
increasing or decreasing the dielectric constant the height of the energy
transfer spectra decreases or increases, respectively. This is because the
the energy transfer is inversely proportional to the square of the
dielectric constant, as shown in Eq. (24). This effect has also been
verified experimentally by Dong et al.,$^{32}$ where upon integrating a
molecular beacon to a CdTe-QD it was found that the fluorescence quenching
due to graphene is modified. We also note that at certain values of probe
detuning, say for example $\delta _{2}\approx \pm 1.5$ $\mu $eV, the
sensitivity of the energy transfer rate to the change in dielectric constant
is quite high. This is an interesting finding, particularly if one considers
that the present hybrid system can be used to fabricate nano-biosensors.%
\FRAME{ftbpFU}{3.3927in}{4.4053in}{0pt}{\Qcb{(a) Energy transfer rate from
the QD to graphene when the dielectric function of the QD is taken as $%
\protect\epsilon _{d}=10$ (dotted curve), $12$ (solid curve) and $14$
(dashed curve). Here $R=13$ nm and $\protect\delta _{3}=0$. (b) Energy
transfer rate from the QD to graphene when the thickness of graphene\ is
varied between one layer (solid curve) or two (dashed curve). Here $R=13$ nm
and $\protect\delta _{3}=0$. Inset: Polarization of the probe and control
fields.}}{}{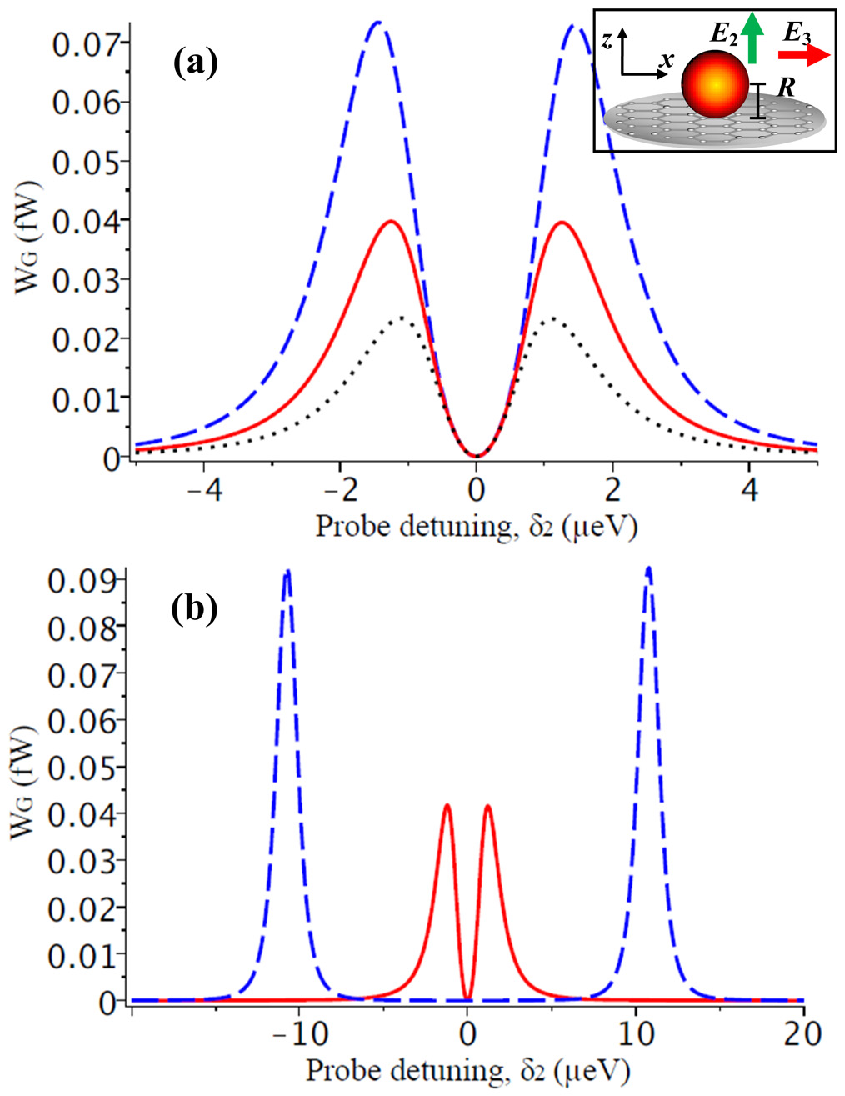}{\special{language "Scientific Word";type
"GRAPHIC";maintain-aspect-ratio TRUE;display "USEDEF";valid_file "F";width
3.3927in;height 4.4053in;depth 0pt;original-width 3.346in;original-height
4.3543in;cropleft "0";croptop "1";cropright "1";cropbottom "0";filename
'figure5.eps';file-properties "XNPEU";}}

We have also studied the effect of the number of graphene layers on the
energy transfer\ spectrum. In Fig. 5b we have plotted the energy transfer
rate to graphene when a single graphene layer or two layers are considered.
Here the ratio $L_{x}/L_{z}=20$ is preserved in order to keep the surface
plasmon polariton resonance frequency constant. Note that for two layers of
graphene the height of the energy transfer peak increases. As we add
additional layers of graphene, we increase its volume. In turn, the DDI
between the QD and graphene is enhanced. Therefore, both the height of the
peaks in the energy transfer spectrum and their splitting are increased.
This effect has also been verified experimentally by Chen et al.$^{31}$,
where it was found that increasing the number of graphene layers in a
CdSe/ZnS nanocrystal-graphene composite system enhanced the QD fluorescence
quenching effect.

We now consider an alternative configuration for the QD-graphene
nanocomposite system where both transitions $\left\vert 2\right\rangle
\leftrightarrow \left\vert 1\right\rangle $ and $\left\vert 3\right\rangle
\leftrightarrow \left\vert 1\right\rangle $ couple with the surface plasmons
in the graphene nanodisk. In this configuration we consider that both $%
\omega _{12}$ and $\omega _{13}$ are close to $\omega _{sp}^{x}$ and the
transition dipole moments (fields) $\mu _{12}$ (probe field) and $\mu _{13}$
(control field) are aligned along the $x$- and $y$-directions, respectively.
In Fig. 6a we have plotted the energy transfer rate from the QD to graphene\
as a function of the probe field detuning while varying $R$. Here the
physical parameters are the same as considered in our previous calculations.
Note that we see two peaks due to the DDI effect as in the first
configuration. Previously the two peaks were symmetric, but in this case
they are asymmetric. This is due to the self-induced DDI parameter $\Lambda
_{2}$, which causes both peaks to shift towards positive detuning due to the
change in the effective probe detuning from $\delta _{2}$ to $\delta _{2}+$ $%
\Delta _{2d}$, as shown in Eq. (\ref{rho_12}). Here also the width of both
peaks increases due to the non-radiative decay $\Gamma _{2d}$. In the
previous configuration the self-induced DDI parameter $\Lambda _{2}$ was
zero because there was no coupling between the QD\ transition $\left\vert
2\right\rangle \leftrightarrow \left\vert 1\right\rangle $ and graphene.\
These effects are enhanced by decreasing $R$.\FRAME{ftbpFU}{3.3927in}{%
4.4763in}{0pt}{\Qcb{Energy transfer rate from the QD to graphene as a
function of probe detuning $\protect\delta _{2}$ for the second
configuration of dipole moments and fields (see inset). (a) The QD-graphene
separation is varied from $R=13$ nm (solid curve) to $R=11$ nm (dashed
curve). (b) $R=11$ nm and the lower band edge of the photonic crystal is
taken as $\protect\varepsilon _{v}=\hbar \protect\omega _{12}+17.88$ meV
(solid curve) and $\protect\varepsilon _{v}=\hbar \protect\omega _{12}+0.56$
meV (dashed curve). Here $\protect\delta _{3}=0$. Inset: Polarization of the
probe and control fields.}}{}{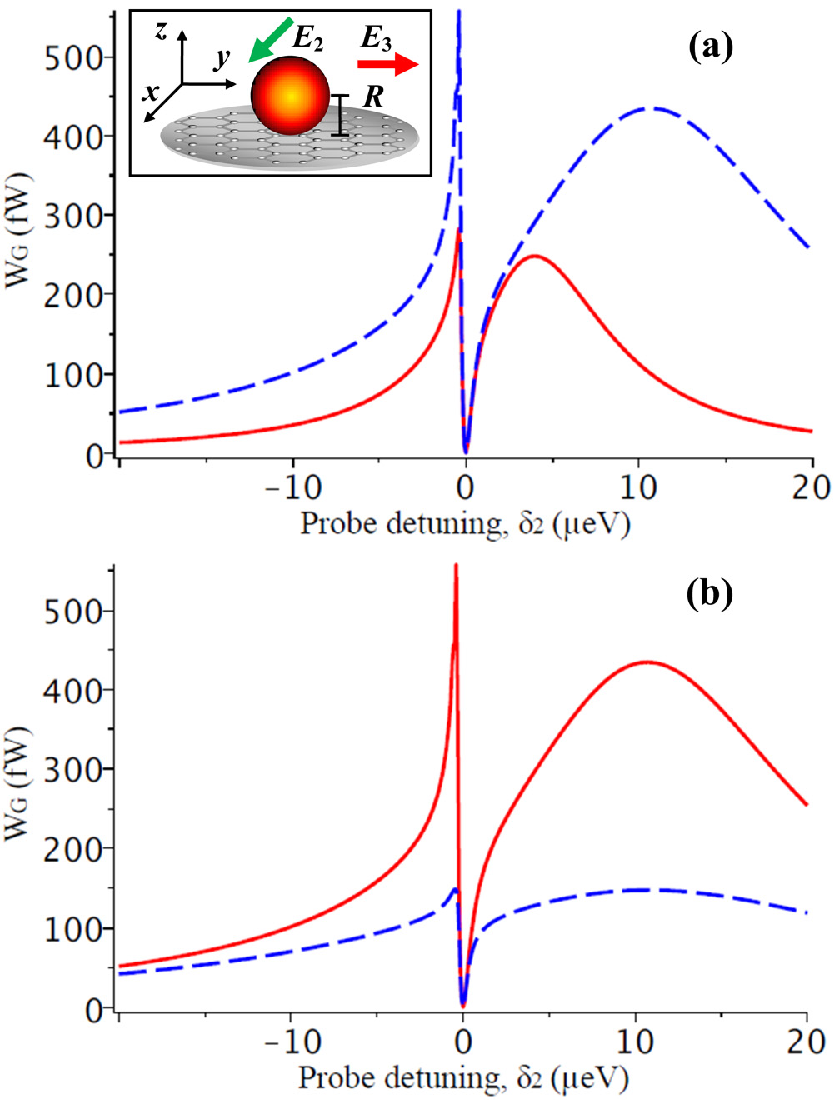}{\special{language "Scientific
Word";type "GRAPHIC";maintain-aspect-ratio TRUE;display "USEDEF";valid_file
"F";width 3.3927in;height 4.4763in;depth 0pt;original-width
3.346in;original-height 4.4244in;cropleft "0";croptop "1";cropright
"1";cropbottom "0";filename 'figure6.eps';file-properties "XNPEU";}}

In Fig. 6b the effect of the photonic crystal is investigated in the same
way as in Fig. 3, and similar results are found as for the previous
configuration. By applying an external pump laser field to the polystyrene
photonic crystal, the power transfer to graphene can be switched from high
to low values. We note that due to the asymmetry of the power transfer
spectrum in this configuration, the sensitivity of this switching effect can
change drastically depending on the value of probe field detuning. For
example, negative detunings close to $\delta _{2}=0$ show a sharp peak in
the energy transfer spectrum when the pump field is absent, and this peak is
suppressed when the pump field is applied.\FRAME{ftbpFU}{3.3918in}{1.9908in}{%
0pt}{\Qcb{Energy absorption rate of to the ladder-type QD\ as a function of
probe field detuning $\protect\delta _{2}$ when the QD-graphene nanodisk\
separation $R$ is varied. Here $\hbar \protect\omega _{23}=0.8036$ eV and
the intensities of the probe and control fields are $1.0$ and $3.0$ W/cm$%
^{2} $, respectively. Other parameters are the same as considered
previously. Inset: Schematic of the QD-graphene hybrid system with
ladder-type energy level structure. Here DDI coupling occurs only for the $%
\left\vert 2\right\rangle \leftrightarrow \left\vert 3\right\rangle $
transition.}}{}{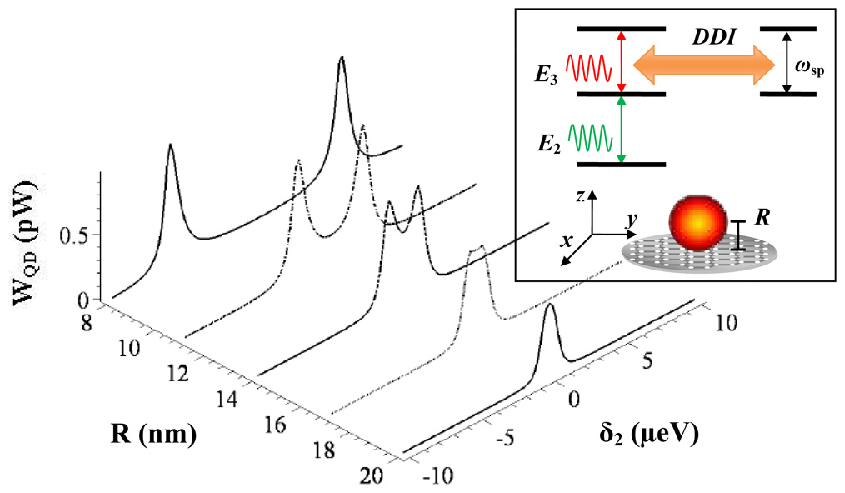}{\special{language "Scientific Word";type
"GRAPHIC";maintain-aspect-ratio TRUE;display "USEDEF";valid_file "F";width
3.3918in;height 1.9908in;depth 0pt;original-width 3.3451in;original-height
1.9519in;cropleft "0";croptop "1";cropright "1";cropbottom "0";filename
'figure7.eps';file-properties "XNPEU";}}

Finally, we investigate the energy absorption rate in a ladder-type QD\
coupled with the graphene nanodisk. The formulation for this system is given
in the appendix. We consider the case where the control field is coupled
with the QD\ transition $\left\vert 2\right\rangle \leftrightarrow
\left\vert 3\right\rangle $ and the graphene nanodisk, while the probe field
is only coupled to the QD transition $\left\vert 1\right\rangle
\leftrightarrow \left\vert 2\right\rangle $. Here the resonance frequency $%
\omega _{23}$ lies near $\omega _{sp}^{x}$, while $\omega _{12}$ is
uncoupled from both $\omega _{sp}^{x}$ and $\omega _{sp}^{z}$. This
situation is analogous to that explored in Fig. 2 for the lambda-type QD. In
Fig. 7 the energy absorption rate in the ladder-type QD is plotted when the
QD-graphene nanodisk separation $R$ is varied. We find that the power
absorption spectrum gives two peaks and a minimum when $R$ is small (i.e. $%
R=8$ nm), as was found in Fig. 2. Note that for the ladder-type QD, the
narrow minima due to electromagnetically induced transparency does not
appear. This is because the same electromagnetically induced transparency
effect does not appear in ladder-type systems.$^{39}$ We have also
investigated the effect of the photonic crystal on the energy absorption
rate in the ladder-type QD, and the results are shown in Fig. 8. Again we
consider that a pump field of intensity $31.0$ GW/cm$^{2}$ is applied, which
causes the photonic crystal band edge to shift and increases the decay rate
of the QD. Note that the power absorption spectrum merges into a broad peak
in the same way as it did for the lambda-type QD (see Fig. 3b). Here,
however, the narrow minima present in the lambda-type QD is absent and we
clearly see two peaks merging into one.\FRAME{ftbpFU}{3.3918in}{3.3771in}{0pt%
}{\Qcb{Energy absorption rate of the ladder-type QD\ as a function of probe
field detuning $\protect\delta _{2}$ when the lower band edge of the
photonic crystal is taken as $\protect\varepsilon _{v}=\hbar \protect\omega %
_{23}+17.88$ meV (a) and $\protect\varepsilon _{v}=\hbar \protect\omega %
_{23}+0.10$ meV (b). Here $R=10$ nm, $\protect\delta _{3}=0$ and other
parameters are the same as in Fig. 7. Inset: Schematic of the QD-graphene
hybrid system with ladder-type energy level structure. Here DDI coupling
occurs only for the $\left\vert 2\right\rangle \leftrightarrow \left\vert
3\right\rangle $ transition.}}{}{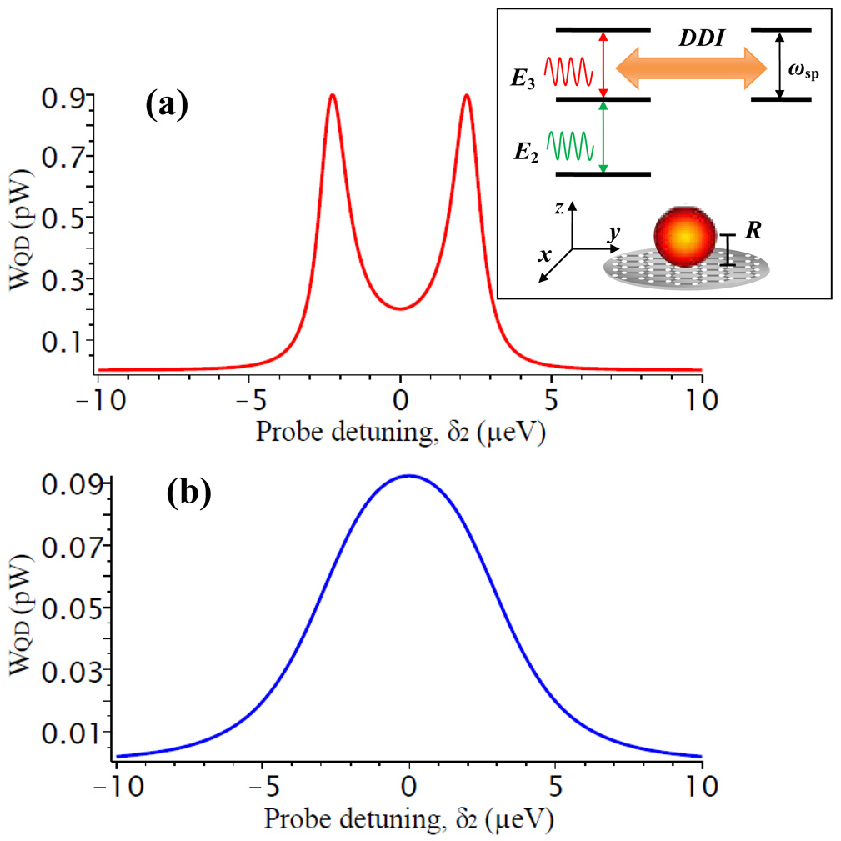}{\special{language "Scientific
Word";type "GRAPHIC";maintain-aspect-ratio TRUE;display "USEDEF";valid_file
"F";width 3.3918in;height 3.3771in;depth 0pt;original-width
3.3451in;original-height 3.3304in;cropleft "0";croptop "1";cropright
"1";cropbottom "0";filename 'figure8.eps';file-properties "XNPEU";}}

\section{Conclusions}

We have investigated the dipole-dipole interaction (DDI) and energy transfer
in a quantum dot (QD)-graphene nanodisk system embedded within a photonic
crystal. Our simulations predict that in this system, multiple excitonic
states (dressed states) can be created in the quantum dot and then
transferred to graphene with different frequencies. This phenomenon occurs
purely due to the DDI between the QD and graphene, and results in energy
transfer. We have demonstrated that the energy absorption of the QD\ and/or
the energy transfer from the QD to graphene can be switched on and off by
changing the strength of the DDI or by applying an intense external field to
the photonic crystal. We have also verified our findings qualitatively with
recent experimental data on energy transfer in QD-graphene nanocomposite
systems. Our numerical results provide motivation for future experimental
and theoretical investigations on nanocomposites made from graphene, carbon
nanotubes, quantum dots and photonic crystals. The present theory can be
applied to hybrid systems consisting of graphene with quantum emitters such
as quantum dots, nanocrystals, atoms and chemical or biological molecules;
the only requirement is the quantum emitter should have at least three
states. The proposed nanocomposite system can be used to fabricate
nano-sensors, all optical nano-switches, energy transfer devices and energy
storage devices.

\section{Appendix}

We consider a three-level quantum dot in the ladder configuration, where $%
\left\vert 1\right\rangle $, $\left\vert 2\right\rangle $ and $\left\vert
3\right\rangle $ denote the ground, first excited and second excited states,
respectively. The probe field with amplitude $E_{2}$ and frequency $\omega
_{2}$ is coupled between states $\left\vert 1\right\rangle $ and $\left\vert
2\right\rangle $, while the control field with amplitude $E_{3}$ and
frequency $\omega _{3}$ is coupled between states $\left\vert 2\right\rangle 
$ and $\left\vert 3\right\rangle $. The decay of level $\left\vert
2\right\rangle $ to level $\left\vert 1\right\rangle $ (level $\left\vert
3\right\rangle $ to level $\left\vert 2\right\rangle $) is given as $\Gamma
_{21}$ ($\Gamma _{32}$). Using the same methods as for the lambda-type
system, the density matrix equations of motion for the ladder-type energy
level configuration are then obtained as%
\begin{eqnarray}
\frac{d\rho _{22}}{dt} &=&-\Gamma _{21}\rho _{22}+\Gamma _{32}\rho
_{33}+iR_{2}e^{-i\theta _{2}}\rho _{21}+i\Lambda _{2}^{\ast }\rho _{12}\rho
_{21}  \TCItag{A1} \\
&&-iR_{2}e^{i\theta _{2}}\rho _{12}-i\Lambda _{2}\rho _{21}\rho
_{12}-iR_{3}e^{-i\theta _{3}}\rho _{32}  \notag \\
&&-i\Lambda _{3}^{\ast }\rho _{23}\widetilde{\rho }_{32}+iR_{3}e^{i\theta
_{3}}\rho _{23}+i\Lambda _{3}\rho _{32}\rho _{23}  \notag \\
\frac{d\rho _{33}}{dt} &=&-\Gamma _{32}\rho _{33}+iR_{3}e^{-i\theta
_{3}}\rho _{32}+i\Lambda _{3}^{\ast }\rho _{23}\rho _{32}  \TCItag{A2} \\
&&-iR_{3}e^{i\theta _{3}}\rho _{23}-i\Lambda _{3}\rho _{32}\rho _{23}  \notag
\\
\frac{d\rho _{21}}{dt} &=&d_{21}\rho _{21}+iR_{2}e^{i\theta _{2}}\left( \rho
_{22}-\rho _{11}\right) -iR_{3}e^{-i\theta _{3}}\rho _{31}  \TCItag{A3} \\
&&-i\Lambda _{3}^{\ast }\rho _{23}\rho _{31}  \notag \\
\frac{d\rho _{32}}{dt} &=&d_{32}\rho _{32}+iR_{3}e^{i\theta _{3}}\left( \rho
_{33}-\rho _{22}\right) +iR_{2}e^{-i\theta _{2}}\rho _{31}  \TCItag{A4} \\
&&+i\Lambda _{2}^{\ast }\rho _{12}\rho _{31}  \notag \\
\frac{d\rho _{31}}{dt} &=&\left( i\delta _{2}+i\delta _{3}-\Gamma
_{32}\right) \rho _{31}-iR_{3}e^{i\theta _{3}}\rho _{21}-i\Lambda _{3}\rho
_{32}\rho _{21}  \TCItag{A5} \\
&&+iR_{2}e^{i\theta _{2}}\rho _{32}+i\Lambda _{2}\rho _{21}\rho _{32}  \notag
\end{eqnarray}%
where%
\begin{eqnarray*}
d_{21} &=&i\delta _{2}+i\Delta _{2d}-\Gamma _{21}-\Gamma _{2d} \\
d_{32} &=&i\delta _{3}+i\Delta _{3d}-\Gamma _{32}-\Gamma _{3d}
\end{eqnarray*}%
and%
\begin{eqnarray*}
\Delta _{2d} &=&\func{Re}\left( \Lambda _{2}\right) \left( \rho _{22}-\rho
_{11}\right) \\
\Gamma _{2d} &=&\func{Im}\left( \Lambda _{2}\right) \left( \rho _{22}-\rho
_{11}\right) \\
\Delta _{3d} &=&\func{Re}\left( \Lambda _{3}\right) \left( \rho _{33}-\rho
_{22}\right) \\
\Gamma _{3d} &=&\func{Im}\left( \Lambda _{3}\right) \left( \rho _{33}-\rho
_{22}\right)
\end{eqnarray*}%
In the above expressions, all quantities are the same as given previously
for the lambda-type system but with the substitutions $\omega
_{13}\rightarrow \omega _{23}$, $\mu _{13}\rightarrow \mu _{23}$ and $\Gamma
_{31}\rightarrow \Gamma _{32}$. The QD energy absorption rate and the power
transfer in this system are calculated using Eqs. (23) and (24) with the
subsitutions $\rho _{11}\rightarrow \rho _{ii}$, $\rho _{13}\rightarrow \rho
_{23}$.


\begin{thebibliography}{99}
\bibitem{1} M. Achermann, J. Phys. Chem. Lett. 1, 2837 (2010).

\bibitem{2} R. D. Artuso and G. W. Bryant, Phys. Rev. B 82, 195419 (2010).

\bibitem{3} S. M. Sadeghi, L. Deng, X. Li, and W.-P. Huang, Nanotechnology
20, 365401 (2009).

\bibitem{4} M.-T. Cheng, S.-D. Liu, H.-J. Zhou, Z.-H. Hao, and Q.-Q. Wang,
Opt. Lett. 32, 2125 (2007).

\bibitem{5} M. Durach, A. Rusina, V. I. Klimov, and M. I. Stockman, New J.
Phys. 10, 105011 (2008).

\bibitem{6} J. M. Luther, P. K. Jain, T. Ewers, and A. P. Alivisatos, Nat.
Mater. 10, 361 (2011).

\bibitem{7} J. D. Joannopoulos, S. G. Johnson, J. N. Winn, and R. D. Meade,
Photonic Crystals: Molding the Flow of Light, 2nd ed. (Princeton University
Press, Princeton, New Jersey, 2008).

\bibitem{8} A. Chutinan, S. John, and O. Toader, Phys. Rev. Lett. 90, 123901
(2003).

\bibitem{9} Y. Liu, F. Qin, Z.-Y. Wei, Q.-B. Meng, D.-Z. Zhang, and Z.-Y.
Li, Appl. Phys. Lett. 95, 131116 (2009).

\bibitem{10} F. Qin, Y. Liu, and Z.-Y. Li, J. Opt. 12, 035209 (2010).

\bibitem{11} J. D. Cox and M. R. Singh, J. Appl. Phys. 108, 083102 (2010).

\bibitem{12} N. T\'{e}treault, E. Arsenault, L.-P. Heiniger, N. Soheilnia,
J. Brillet, T. Moehl, S. Zakeeruddin, G. A. Ozin, and M. Gratzel, Nano Lett.
11, 4579 (2011).

\bibitem{13} C. E. Talley, J. B. Jackson, C. Oubre, N. K. Grady, C. W.
Hollars, S. M. Lane, T. R. Huser, P. Nordlander, and N. J. Halas, Nano Lett.
5, 1569 (2005).

\bibitem{14} N. I. Zheludev, Opt. Photon. News, 22, 30 (2011).

\bibitem{15} H. A. Atwater and A. Polman, Nat. Mater. 9, 205 (2010).

\bibitem{16} L. Novotny and N. Van Hulst, Nat. Photon. 5, 83 (2011).

\bibitem{17} A. Gonzalez-Tudela, D. Martin-Cano, E. Moreno, L.
Martin-Moreno, C. Tejedor, and F. J. Garcia-Vidal, Phys. Rev. Lett. 106,
020501 (2011).

\bibitem{18} P. R. West, S. Ishii, G. V. Naik, N. K. Emani, V. M. Shalaev,
and A. Boltasseva, Laser Photonics Rev. 4, 795 (2010).

\bibitem{19} A. H. Castro Neto, F. Guinea, N. M. R. Peres, K. S. Novoselov,
and A. K. Geim, Rev. Mod. Phys. 81, 109 (2009).

\bibitem{20} F. Bonaccorso, Z. Sun, T. Hasan, and A. C. Ferrari, Nat.
Photon. 4, 611 (2010).

\bibitem{21} F. Schedin, E. Lidorikis, A. Lombardo, V. G. Kravets, A. K.
Geim, A. N. Grigorenko, K. S. Novoselov, and A. C. Ferrari, ACS Nano 4, 5617
(2010).

\bibitem{22} F. H. L. Koppens, D. E. Chang, and F. J. G. Abajo, Nano Lett.
11, 3370 (2011).

\bibitem{23} K. P. Loh, Q. Bao, G. Eda, and M. Chhowalla, Nat. Chem. 2, 1015
(2010).

\bibitem{24} R. R. Nair, P. Blake, A. N. Grigorenko, K. S. Novoselov, T. J.
Booth, T. Stauber, N. M. R. Peres, and A. K. Geim, Science 320, 1308 (2008).

\bibitem{25} T. Mueller, F. Xia, and P. Avouris, Nat. Photon. 4, 297 (2010).

\bibitem{26} M. Jablan, H. Buljan, and M. Soljacic, Phys. Rev. B 80, 245435
(2009).

\bibitem{27} Q. Bao, H. Zhang, B. Wang, Z. Ni, C. H. Y. X. Lim, Y. Wang, D.
Y. Tang, and K. P. Loh, Nat. Photon. 5, 411 (2011).

\bibitem{28} C. Tegenkamp, H. Pfnur, T. Langer, J. Baringhaus, and H. W.
Schumacher, J. Phys.: Condens. Matter 23, 012001 (2011).

\bibitem{29} A. Manjavacas, P. Norlander, and F. J. G. Abajo, ACS Nano 6,
1724 (2012).

\bibitem{30} A. Cao, Z. Liu, S. Chu, M. Wu, Z. Ye, Z. Cai, Y. Chang, S.
Wang, Q. Gong, and Y. Liu, Adv. Mater. 22, 103 (2010).

\bibitem{31} Z. Chen, S. Berciaud, C. Nuckolls, T. F. Heinz, and L. E. Brus,
ACS Nano 4, 2964 (2010).

\bibitem{32} H. Dong, W. Gao, F. Yan, H. Ji, and H. Ju, Anal. Chem. 82, 5511
(2010).

\bibitem{33} P. Wang, T. Jiang, C. Zhu, Y. Zhai, D. Wang, and S. Dong, Nano
Res. 3, 794 (2010).

\bibitem{34} E. Shafran, B. D. Mangum, and J. M. Gerton, Nano Lett. 10, 4049
(2010).

\bibitem{35} G. Goncalves, P. A. A. P. Marques, C. M. Granadeiro, H. I. S.
Nogueira, M. K. Singh, and J. Gracio, Chem. Mater. 21, 4796 (2009).

\bibitem{36} C. Xu, X. Wang, and J. Zhu, J. Phys. Chem. C 112, 19841 (2008).

\bibitem{37} R. Patakfalvi, D. Diaz, P. Santiago-Jacinto, G.
Rodriguez-Gattorno, and R. Sato-Berru, J. Phys. Chem. C 111, 5331 (2007).

\bibitem{38} S. Evangelou, V. Yannopapas, and E. Paspalakis, Phys. Rev. A
83, 023819 (2011); Phys. Rev. A 83, 055805 (2011); V. Yannopapas, E.
Paspalakis, and N. V. Vitanov, Phys. Rev. Lett. 103, 063602 (2009).

\bibitem{39} M. O. Scully and M. S. Zubairy, Quantum optics (Cambridge
University Press, Cambridge, UK, 1997).

\bibitem{40} M. Fleischhauer, A. Imamoglu, and J. P. Marangos, Rev. Mod.
Phys. 77, 633 (2005).

\bibitem{41} M. V. Gurudev Dutt, J. Cheng, B. Li, X. Xu, X. Li, P. R.
Berman, D. G. Steel, A. S. Bracker, D. Gammon, S. E. Economou, R.-B. Liu,
and L. J. Sham, Phys. Rev. Lett. 94, 227403 (2005).

\bibitem{42} X. Xu, B. Sun, P. R. Berman, D. G. Steel, A. S. Bracker, D.
Gammon, and L. J. Sham, Nat. Phys. 4, 692 (2008).

\bibitem{43} J. M. Elzerman, K. M. Weiss, J. Miguel-Sanchez, and A.
Imamoglu, Phys. Rev. Lett. 107, 017401 (2010).

\bibitem{44} D. Sarid and W. A. Challener, Modern Introduction to Surface
Plasmons: Theory, Mathematica Modeling, and Applications (Cambridge
University Press, New York, 2010).

\bibitem{45} P. Hanarp, M. K\"{a}ll, and D. S. Sutherland, J. Phys. Chem. B
107, 5768 (2003).

\bibitem{46} C. Langhammer, M. Schwind, B. Kasemo, and I. Zori\'{c}, Nano
Lett. 8, 1461 (2008).

\bibitem{47} C. F. Bohren and D. R. Huffman, Absorption and scattering of
light by small particles (Wiley-Interscience, New York, 1983), Chap. 5.

\bibitem{48} E. C. Le Ru, Principles of Surface-Enhanced Raman Spectroscopy
and Related Plasmonic Effects (Elsevier, Amsterdam, 2009), Appendix G.

\bibitem{49} B. E. Kane, Phys. Rev. B 82, 115441 (2010).

\bibitem{50} S. John and J. Wang, Phys. Rev. B 43, 12772 (1991).

\bibitem{51} I. Haque and M. R. Singh, J. Phys.: Condens. Matter 19, 156229
(2007).
\end{thebibliography}
\end{document}